\documentclass[aps,prl,preprint,groupedaddress,nofootinbib]{revtex4-1}
\usepackage{graphicx}
\begin{document}


\title{\large  
Matter-Dark Matter Coincidence  and Mirror World
}


\author{\bf Rabindra N. Mohapatra$^a$}
\author{Nobuchika Okada$^b$}
\affiliation{}
\affiliation{$^a$ Maryland Center for Fundamental Physics and Department of Physics, University of Maryland, College Park, Maryland 20742, USA}
\affiliation{$^b$ Department of Physics, University of Alabama, Tuscaloosa, Alabama 35487, USA}


\date{\today}

\begin{abstract} Why matter and dark matter contents of the universe are of the same order of magnitude, is one of the puzzles of modern cosmology. At the face of it, this would seem to point towards a basic similarity between matter and dark matter, suggesting perhaps the widely discussed mirror world picture as an ideal setting for a discussion of this issue. Here we outline a new and simple mirror world scenario to explain this puzzle. Our model  uses Affleck-Dine mechanism to generate baryon asymmetry and dark matter relic density leading to an asymmetric dark matter picture. We find that for a certain parameter range of the model, the mirror electron is the unique possibility for dark matter whereas in the complementary parameter range, the mirror baryons constitute the dark matter. In either case, the mirror photon must have  mass in the MeV range for consistency with observations. For the case of mirror electron dark matter, the model predicts a lower bound on the amount of dark radiation i.e. $\Delta N_{eff} \geq 0.007$. 
\end{abstract}

\maketitle

\section{1. Introduction}
One of the puzzles of dark matter cosmology is to understand why the baryon content of the universe is of the same order of magnitude as its dark matter content i.e. $\Omega_{DM }\simeq 5\Omega_B$. We call this matter-dark matter coincidence. This relation could be accidental or there could be some physical reason behind it. If it is the latter, it could reveal some interesting properties about the dark matter as well as our universe.  It is therefore worth investigating theories where that happens.

When one considers dark matter physics as being unrelated to matter physics described by the standard model, as is done often, this relation between baryon excess and dark matter content of the universe would appear to be accidental. Supersymmetric WIMP or the axion dark matter pictures are examples of this. On the  other hand, the above relation could be suggesting that at some basic level, there is a similarity between matter and dark matter,  providing a basis for this coincidence. A scenario where the matter-dark matter similarity is evident is the mirror world picture where the known standard model is believed to co-exist with its parity duplicate sector of matter and forces, the mirror sector~\cite{LY, KOZ, M1, M2, M3, M4, M5, M6, M7} interacting with SM only via gravity or similar ultra-weak forces. Departures from exact parity (or $Z_2$) symmetry could arise due to spontaneous or soft breaking of the symmetry. We assume in our discussion that this symmetry breaking leads to the weak scale in the mirror sector being higher than that of the visible sector, This is the so-called asymmetric mirror model. The lightest particle of the mirror world would then have all the characteristics of a dark matter and could indeed be the dark matter of the universe if its abundance is appropriate. The mirror world picture would thus appear to be an ideal setting for addressing the puzzle of matter-dark matter coincidence. 

As is well known, the consistency of a general mirror world picture with the success of big bang nucleosynthesis (BBN) requires that the mirror sector of the universe be cooler than the visible sector~\cite{Kolb, Hodges, BDM}  to suppress the contribution of mirror neutrinos and mirror photon to BBN. This is known as asymmetric reheating and every mirror world scenario must have a mechanism for asymmetric reheating. There are various ways to achieve that. One can take two inflatons, one for the visible sector and another for the mirror sector, and choose different histories such that it makes the reheating temperature different in the two sectors (see for instance ~\cite{Vilenkin, cline}). One could also arrange the neutrino sector for this purpose, see~\cite{chacko}. Another way to achieve this goal~\cite{BDM} would be to choose the inflaton to be a mirror $Z_2$ odd scalar and couple it to the SM and mirror sector fields in such a way that the inflation reheat temperature, $T_R\simeq \sqrt{\Gamma M_P}$ is different in both sectors. In this paper, we adopt a different point of view and assume a particularly simple spectrum  for gauge singlet fermions (denoted by $N, S$ and $N', S'$) in such a way  as to guarantee asymmetric reheating via the inflaton decay. We use soft breaking of mirror $Z_2$ symmetry for the singlet neutrinos  in such a way as to guarantee that the inflaton partial decay widths to mirror particles will be less than that to SM particles, ensuring that mirror sector reheat temperature is lower than the visible sector. We will assume lepton numbers to be exact in the visible as well as the mirror sector so that the familiar neutrino is a Dirac fermion.

Within this picture, we can assume that just like in the visible sector, in the mirror sector, there is a mirror matter-mirror antimatter asymmetry and the excess mirror matter is the dark matter as in the asymmetric dark matter scenario~\cite{shmuel, sekhar, kaplan}(for reviews, see~\cite{DM, PV, zurek}), which is an alternative way to generate the relic density of dark matter. 
The question then arises as to which particle of the mirror sector is  the dark matter, out of many possibilities such as mirror proton or neutron, mirror atoms, mirror electron, mirror neutrinos, mirror photon etc.  Pursuing the mirror world alternative for dark matter as a platform to answer the coincidence puzzle alluded to above, we find that for a particular range of parameters, the mirror electron  turns out to be the unique choice for dark matter whereas in the complementary parameter range, it is the mirror proton.  The identification of the dark matter particle and the scenario that leads to it are the new results of this paper. 

There have been other mirror world proposals for resolution of the baryon-dark matter  puzzle, see ref.~\cite{An, cui, nath, farina, FMS, Raman, volkas, ibe, curtin, burdman, murgui, yi, borah}. Our scenario is different in a fundamental way from these as well as other recent speculations in this regard~\cite{Hook, Abhi}. 

Our basic strategy is to use the Affleck-Dine~\cite{AD} (AD) mechanism for generating the mirror matter-mirror anti-matter as well as visible sector matter-anti-matter asymmetry in the asymmetric mirror world scenario. This provides the first key step towards resolving the baryon-dark matter coincidence puzzle. We assume that  the AD field responsible for generation of matter asymmetry is also the inflaton field. Furthermore, we incorporate the new way to generate asymmetric reheating from the inflaton decay into our picture.  Affleck-Dine mechanism then leads to asymmetry in the lepton number in both sectors, which leads to constraints on the parameters of the model. In one parameter range, our scenario implies that the mirror sphalerons go out of equilibrium before inflaton decay so that mirror lepton asymmetry generated from singlet mirror fermion decay, does not get a chance to be converted to mirror baryons. In this case therefore  the lightest mirror lepton, the mirror electron, with a mass of few GeV, remains as the asymmetric dark matter of the universe and provides a resolution of the matter-dark matter coincidence puzzle. On the other hand, in a complementary parameter range, the mirror sphaleron remains active after inflaton decay so that the mirror baryons are generated from the mirror lepton asymmetry and become the dark matter. The mirror baryons and mirror anti-baryons as well as the mirror leptons and anti-leptons annihilate each other due to their "stronger" force leaving a negligible symmetric content for dark matter in both cases. 

 Since mirror electron has mirror electric charge, dark matter is of the self-interacting variety (SIDM) ~\cite{spergel, haibo} due to mirror electromagnetic interaction. Even though the self interaction force of the dark matter is repulsive, if the mirror photon has mass, this can still be an acceptable dark matter as shown in ref.~\cite{haibo2}.  If the model has photon-mirror photon mixing, it can also make the dark matter weakly visible, but we do not pursue such an extension here.

The paper is organized as follows: in sec.~2, we present the mirror model used in the paper;  in sec.~3, we discuss how Dirac neutrino mass arises in our model planting the seed for asymmetric reheating, which is described in sec.~4; sec.~5 describes the cosmological evolution and Affleck-Dine matter-dark matter generation and how baryon-dark matter coincidence puzzle is resolved in the model; in sec. 6, we describe the constraints on the model from baryon dark matter correspondence. In sec. 7, we present other constraints on the model due to wash-out effects as well as our prediction for $\Delta N_{eff}$ due to mirror dark radiation. In sec. 8, we discuss the necessity of a massive mirror photon in our model to fit observations.  In sec. 9, we summarize our results.

\section{2. The model} The mirror world picture consists of the standard model of particles and forces coexisting with its parity (or $Z_2$-symmetric) duplicate. The basic particle content of the model is given in table I~\cite{LY, KOZ, M1, M2, M3, M4, M5, M6, M7}. To this, we may add other particles which respect the basic philosophy of mirror models and the $Z_2$ symmetry. The $Z_2$ symmetry guarantees that the couplings in the SM and the mirror sectors are same, keeping the number of couplings same as in the visible sector. We also entertain soft $Z_2$ breakings for the scalar sector so that the scalar masses as well as symmetry breakings in the two sectors could be different. In the table I, we display all the SM particles and their mirror partners  including the extra singlet fermions (RHN) $N, S$ and $N', S'$ as well as scalars $\eta$ and $\eta'$ ( iso-singlet partners with charges $+1$), the latter being the new additions to the model for our purpose, as we discuss below.  We also have a complex scalar field $\Phi$, which is  $Z_2$-even, and is common to both sectors. As we will explain, $\Phi$ plays the key role in generating baryon and dark matter excess and causing inflationary expansion of the universe. We will give  mirror, $\eta^{+ \prime}$,  field a vacuum expectation value (vev)  to make mirror photon massive.

 \begin{table}[h!]
\centering
\begin{tabular}{|c||c||c||c|}\hline
Our world & $SU(3)_c\times SU(2)_L\times U(1)_Y$ & mirror world & $SU(3)'_c \times SU(2)'_R\times U(1)'_{Y}$ \\\hline
{\rm Visible fermions }& &{\rm  mirror fermions} &\\\hline
$Q_L$ & $(3, 2, 1/3)$ &$Q'_R$ & $(3, 2, 1/3)$\\
$u_R$ & $(3, 1,  4/3)$&$u'_L$ & $(3, 1, 4/3)$\\
$d_R$ & $(3, 1,  -2/3)$&$d'_L$ & $(3, 1, -2/3)$\\
$\ell_L$ & $(1, 2,  -1)$&$\ell'_R$ & $(1, 2, -1)$\\
$e_R$ & $(1, 1,  -2)$&$e'_L$ & $(1, 1, -2)$\\\hline
Singlet neutrinos & &Singlet mirror neutrinos&\\\hline
$N_R, S_R$ & $(1,1,0)$ & $N'_L, S'_L$ &  $(1,1,0)$\\\hline
{\rm Gauge bosons} && {\rm Mirror Gauge bosons}&\\\hline
$W, Z, \gamma, {\rm Gluons(G)}$ && $W', Z', \gamma', {\rm G'}$&\\\hline
{\rm Scalar sector} & &{\rm mirror scalar} &\\\hline
$H$ &$(1,2,+1)$&$H'$ & $(1,2,+1)$ \\
$\eta$ & $(1,1,2)$ & $\eta'$ & $(1,1,2)$ \\
$\Phi$ & $(1,1, 0)$ & $\Phi$ & $(1,1, 0)$\\\hline
\end{tabular}
\caption{Gauge quantum numbers of all the fields in the theory. $\Phi$ is a mirror parity even field common to both sectors and it will play the role of inflaton and the Affleck-Dine field to produce the lepton asymmetry. It has $L$ and $L'$ numbers equal to $-2$. Also note the flip from left to right chirality for fermions as we go from SM to mirror sector (and vice versa). The formulae for the standard and mirror electric charges are given by $Q=I_{3L}+\frac{Y}{2}$ and  $Q'=I'_{3R}+\frac{Y'}{2}$ respectively.}
\end{table}

The asymmetric mirror model starts with the following basic parts to the Lagrangian:
The scalar potential describing inflation and the general cosmic evolution is given by:
\begin{eqnarray}
V(\Phi, H, H')~=~&& V_{SM}(H,\eta)~+V_{SM'}(H', \eta')+M^2_\Phi |\Phi|^2+\lambda_\Phi | \Phi|^4  \nonumber\\
&&+(\epsilon m^2_{\Phi}\Phi^2+ h.c. )
+\lambda_{\Phi H}| \Phi|^2 (H^\dagger H+H^{\prime \dagger}H^\prime) , 
\end{eqnarray}
where $$V_{SM}(H,\eta)=-\mu^2_HH^\dagger H +\lambda_H (H^\dagger H)^2+\mu^2_\eta \eta^\dagger\eta+\lambda_\eta (\eta^\dagger\eta)^2+\lambda_{\eta H} \eta^\dagger\eta H^\dagger H,$$
and $$V_{SM}(H',\eta')=-\mu^2_{H'}H^{'\dagger} H' +\lambda_H (H^{'\dagger} H')^2+\mu^2_{\eta'} \eta^{\prime \dagger}\eta^\prime
+\lambda_\eta (\eta^{\prime \dagger}\eta^\prime)^2+\lambda_{\eta H} \eta^{\prime \dagger}\eta^\prime H^{\prime \dagger} H^\prime$$
with $\mu^2_{H^\prime} \gg \mu^2_{H}$. We set all terms in the Higgs potential that connect the visible sector to mirror sector (such as $H^\dagger H H^{\prime \dagger}H^\prime$ etc) to zero to prevent thermalization. We choose $\lambda_{H\Phi}$ to be very small. We have also omitted terms in the potential which play no role in our discussion.

Note that due to different scalar mass terms in the two sectors, the $Z_2$ symmetry is only softly broken. That difference allows for a higher mirror weak scale compared to the familiar SM sector. This also gives the mirror Higgs a mass of order $v'$, much higher than the SM Higgs mass. Similarly, the difference in the $\eta$ masses allows us to give the mirror photon a mass while keeping the familiar photon massless. All dimension four coupling strengths on both sides, however, are same at the tree level as would be required by the $Z_2$ symmetry. 
The Yukawa couplings and singlet lepton masses in the model are:
\begin{eqnarray}
{\cal L}_Y~=~{\cal L}_{Y, SM}+ {\cal L}_{Y, SM'}+y_N(\bar{N}_R\ell_L H+\bar{N'}_L\ell'_R H') +y_\Phi \Phi (N_RN_R+N'_LN'_L) \\\nonumber 
+ M_1 \bar{N}_RN'_L+M_2\bar{N}_RS'_L +M_3\bar{S}_RN'_L+M_4\bar{S}_RS'_L+h.c.
\end{eqnarray}
We assign lepton numbers $ L=L'=1$ to leptonic doublets and the singlet neutrinos $N$ $S$ etc. of the two sectors and $L=L'= -2$ to $\Phi$. This is important for leptogenesis. The Dirac mass term for the singlet neutrinos $M_{a=1,2,3,4}\neq 0$ has the symmetry $L+L'=0$. This gives a Dirac mass to the familiar neutrinos, as we  see below. We choose all the mass parameters $M_a$  to be real for simplicity of analysis. Also we set the couplings of $S, S'$ to $\ell$ and $\ell'$ to zero.

To proceed further with discussion, let us note that the doublet Higgs vevs in the two sectors are 
$ \langle H^0 \rangle =v$ and $ \langle H^{'0} \rangle =v^\prime$ and satisfy $v^\prime \gg v$ as already noted. 
We call the ratio $v^\prime/v\equiv r$ for future notation. 
This is one of the fundamental new parameters of the model with other parameters fixed by known standard model physics and $Z_2$ symmetry.  As we discuss below, the mirror sector will have to have a lower temperature to be compatible with big bang nucleosynthesis; we call this parameter $x= T^\prime_R/T_R$. This will be the second parameter of the model. 

\begin{figure}[tb]
  \centering
 \includegraphics[width=0.7\linewidth]{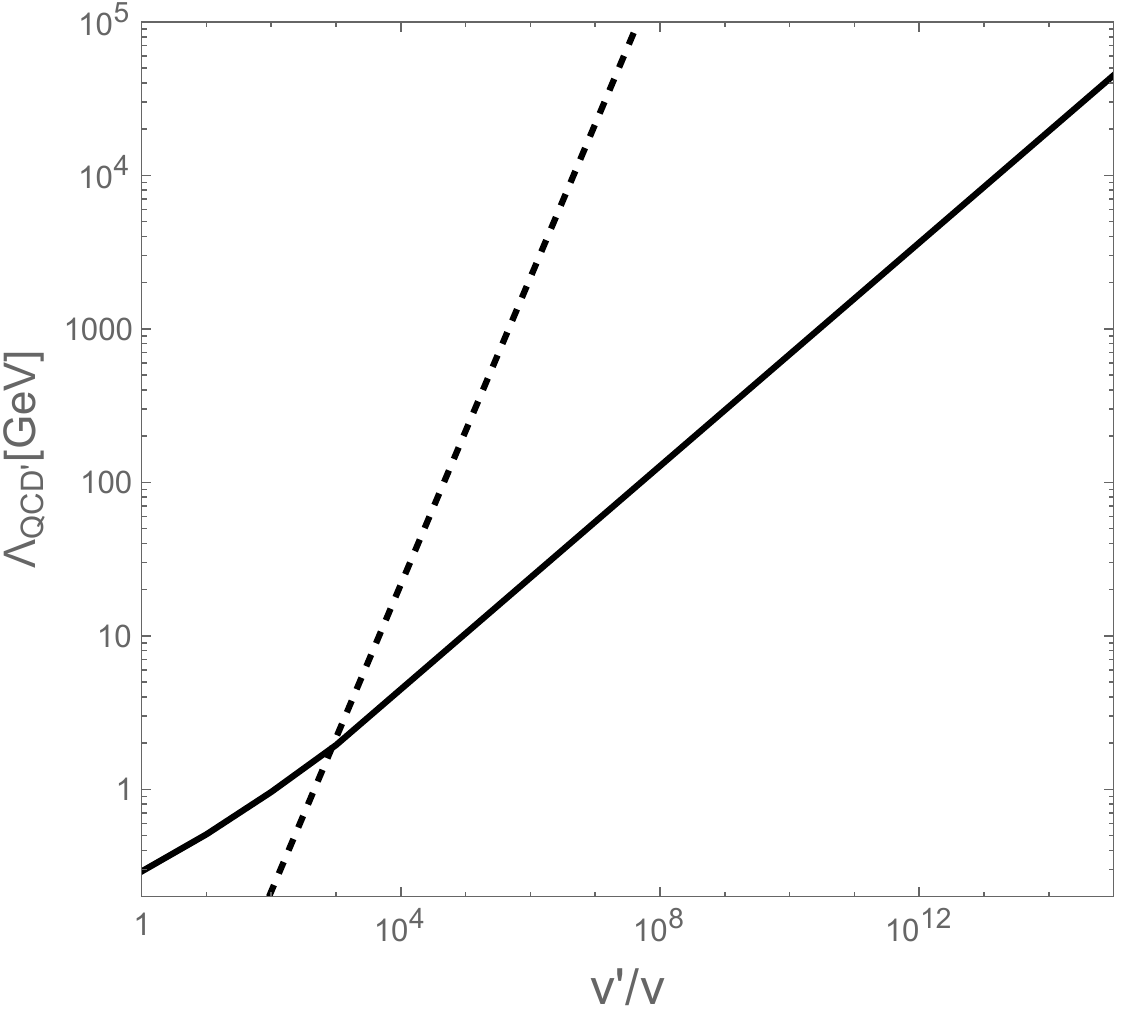}
  \caption{Variation of $\Lambda'_{QCD}$ with $v'/v $ (solid line). The dashed line represents the mass of the mirror up-quark calculated using the ratio $v'/v$. Note that above $v'/v=1000$, the contribution 
to mirror baryon masses will be dominated by the mirror quark current masses.
} 
  \label{fig:1}
  \end{figure}

This asymmetric vev feature of the model has other implications such as for instance, the bare masses of mirror quarks and leptons are $r \equiv v'/v$ times the known quark and lepton masses; it also implies that the strong force between the mirror quarks becomes stronger than known strong force at higher scale i.e. $\Lambda_{QCD'} \gg  \Lambda_{QCD}$.  In Fig.~\ref{fig:1}, we show the mirror QCD scale as a function of $r=v'/v$. Here, we have analyzed the renormalization group equations for QCD and QCD' couplings ($\alpha_{QCD}(\mu)$ and  $\alpha_{QCD'}(\mu)$ ) at the one loop level with the boundary condition that at high scale ($\mu=M_P$)  $\alpha_{QCD}=\alpha_{QCD'}$. $\Lambda_{QCD}$ ($\Lambda_{QCD'}$) are defined as the scales where  $\alpha_{QCD} (\alpha_{QCD'}) =1$.

\section{3. Neutrino mass} 
Before delving into the main details of baryogenesis and dark matter, we discuss the neutrino mass in the model, which is at the root of asymmetric reheating. Due to $L+L'$ symmetry obeyed by the $S, S',N, N'$ mass terms and the lepton Yukawa couplings of the model, the neutrino is a Dirac fermion with $\nu$ and $\nu'$ being the Dirac partners as we see below. 
 The $\nu-\nu'$ and singlet fermion mass matrix is given by 
\begin{eqnarray}
{\cal M}_{\nu,N, S}~=~\begin{array}{ccc} (\bar{\nu_L} & \bar{N'}_L &\bar{S'}_L )\end{array}\left(\begin{array}{ccc} 0 & m&0\\m'& M_1 & M_3\\0 &M_2 & M_4\end{array}\right)\left(\begin{array}{c} \nu'_R\\N_R\\S_R \end{array}\right)
\end{eqnarray}
Each entry in the above matrix is a $3\times 3$ matrix corresponding to three generations of neutrinos, which we suppress. Further, we have, $m=y_Nv$ and $m'=y_Nv'$ with $m'\gg m$. We also assume that 
$M_a \gg m, m'$ and also for simplicity, keep all parameters to be real. We first diagonalize $N-S-N'-S'$ sector:
\begin{eqnarray}
M_{N,S}~=~\begin{array}{cc} ( \bar{N'}_L &\bar{S'}_L )\end{array}\left(\begin{array}{cc}  M_1 & M_3\\ M_2 & M_4\end{array}\right)\left(\begin{array}{c} N_R\\S_R \end{array}\right) .
\end{eqnarray}
This matrix can be diagonalized by bi-orthogonal transformations as follows:
\begin{eqnarray}
\left(\begin{array}{cc}s_\beta & -c_\beta\\c_\beta & s_\beta\end{array}\right) \left(\begin{array}{cc}  M_1 & M_2\\ M_3 & M_4\end{array}\right)\left(\begin{array}{cc}c_\alpha & s_\alpha  
\\-s_\alpha & c_\alpha\end{array}\right) =  \left(\begin{array}{cc}  M & 0\\ 0 & M'\end{array}\right)
\end{eqnarray}
This  leads to the mass eigenstates $N_1=c_\alpha N+ s_\alpha S$ (visible sector) and  $N'_1=s_\beta N'- c_\beta S'$ (mirror sector) forming one Dirac pair with mass $M$ and  $N_2=-s_\alpha N+ c_\alpha S$ (visible sector) and  $N'_2=c_\beta N'+s_\beta S'$ (mirror sector) forming the second pair with mass $M'$.  We choose parameters such that $M'\gg M_\Phi \gg M$ and  $c_{\alpha, \beta} \sim 1$ and $s_{\alpha, \beta} =\delta_{\alpha, \beta} \ll 1$. One can, then, diagonalize the full neutrino mass matrix which leads to the light neutrino masses  given by the approximate eigenvalues of the matrix $m_\nu$ below. The neutrinos are Dirac fermions and they pair up $\nu$ with mirror neutrinos $(\nu'$) to form their mass:
\begin{eqnarray}
m_\nu~=~\frac{y^2_N vv'c_\alpha s_\beta}{M}+\frac{y^2_N vv's_\alpha c_\beta}{M'} ~\simeq ~\frac{y^2_N vv'c_\alpha s_\beta}{M}.
\label{nu_mass}
\end{eqnarray}
We adjust $y_N$ and $M$ so that we get $m_\nu \leq 5\times 10^{-11}$ GeV to fit the scale of the neutrino oscillation data. 
Thus unlike some other mirror models, the mirror neutrinos in our model are very light.

\section{4. Asymmetric reheating} 
One of the essential requirements for mirror models to be viable is that the mirror world must have a lower temperature than the visible world to maintain the success of big bang nucleosynthesis. All the extra relativistic mirror particles e. g. neutrinos, mirror photon etc. contribute less energy density to the Hubble rate. So each mirror model must have a mechanism for generating this asymmetric reheating. 
In our case, the mechanism uses the asymmetry in the $N$ and $N'$ mass sector, which asymmetrizes the  $\Phi$ decay, this being the inflaton field, as we will see below. The reheating in each sector is governed by the decay of the $\Phi$ field to particles in the visible and mirror sectors. We can now rewrite the $\Phi (NN+N'N')$ coupling in the Lagrangian in Eq. (2) as follows, to leading order in the mixings $\alpha$ and $\beta$ in terms of states $N_1$ and $N'_1$. Choosing $c_{\alpha, \beta} \sim 1$ and $s_{\alpha, \beta} =\delta_{\alpha, \beta} \ll 1$), we get
\begin{eqnarray}
{\cal L}_{\Phi N}~=~y_\Phi\Phi ( {N}_{1,R}N_{1,R} + \delta^2_\beta N'_{1,L}N'_{1,L})+ {\rm couplings~ to~ N_2}~+~h.c.
\end{eqnarray}
We have not displayed the couplings to $N_2$ and $N'_2$ since they are chosen to be heavier than the $\Phi$ field and thus do not participate in the reheating process.
Note from the above equation that the $\Phi$ decays to the mirror (prime fields) sector with a suppressed coupling. 
This guarantees that the mirror sector will be cooler due to a lower reheating temperature guaranteed by lower decay rate of the inflaton field $\Phi$.
 Using  $3M^2_P (\Gamma_\Phi)^2\simeq \rho_\Phi\equiv \frac{\pi^2}{30}T^4_R$  (and similarly for the decay to mirror sector), we find that
\begin{eqnarray}
x\equiv \frac{T'_R}{T_R}\simeq ( \delta_\beta) \equiv (Br(\Phi\to N'_1N'_1))^{1/4} ,
\label{x1}
\end{eqnarray}
assuming that the number of degrees of freedom at the  decay epoch  is same in both worlds.
We also choose $\lambda_{\Phi H}\ll 10^{-5}$ so that the $\lambda_{\Phi H}| \Phi|^2 (H^\dagger H+H^{\prime \dagger}H^\prime) $ term does not  bring the mirror sector into equilibrium with the SM sector.

Using Eqs.~(\ref{nu_mass}) and (\ref{x1}), we conclude that
\begin{eqnarray}
m_\nu~\simeq~\frac{y^2_N vv' (Br)^{1/4}}{M}=\frac{y^2_N vv' x}{M}  .
\end{eqnarray}

\section{5. Baryon-Dark matter coincidence} We now come to the main focus of the paper i.e. explaining the matter dark matter coincidence and constraints imposed by this on the parameters of the model. We first outline the general cosmological scenario by which in the early universe, the dark matter content is generated, emphasizing the qualitative aspects and the basic physics in this section. In the following section, we become more quantitative and present the constraints on the model parameters where the idea works. 

To discuss this, we first briefly review the evolution of inflaton and the AD-field $\Phi$~\cite{Stubbs, russian, cline2, herzberg, nobu} in the early universe. Recall that 
\begin{eqnarray}
{\cal L}_\Phi~=~ y_\Phi \Phi(N_R N_R+N'_L N'_L ) + h.c. 
-M^2_\Phi|\Phi|^2 -\lambda_\Phi |\Phi|^4 -\epsilon M^2_\Phi (\Phi\Phi+h.c.), 
\label{Eq:NN_coupling}
\end{eqnarray}
where the $\epsilon$ term breaks lepton number softly by 4 units, leaving a $Z_4$ subgroup of $L$ intact. 
We call  $\Phi$ as the Affleck-Dine (AD) field, which also plays the role of inflaton. 
The dynamics of inflation arises from the non-minimal gravity coupling of the $\Phi$ field given  in \cite{Bezrukov}:
\begin{eqnarray}
{\cal L}_{g}~=~-\frac{1}{2}\int d^4 x  \,\sqrt{-g} \, \left[M^2_P+ 2 \xi |\Phi|^2 \right] R ,
\end{eqnarray}
where $M_P=2.4 \times 10^{18}$ GeV is the reduced Planck mass. 
This Lagrangian is in the Jordan frame, and making a Weyl transformation 
via $g^J_{\mu\nu}\to \left(1+ 2 \xi\frac{|\Phi|^2 }{M^2_P} \right)g^J_{\mu\nu}\equiv g^E_{\mu\nu}$, 
we can go to the Einstein frame, where the potential for $\Phi$ becomes 
\begin{eqnarray}
V^E(\Phi)~=~\frac{V^J(\Phi)}{\left(1+2 \xi\frac{|\Phi|^2}{M^2_P}\right)^2}.
\end{eqnarray}
As we see from the shape of the potential in the above equation, it is flat for $\Phi > M_P/\sqrt{\xi}$, and this is responsible for inflationary expansion 
 of the universe. 
As inflation proceeds, the value of $\Phi$ becomes smaller as it rolls down the potential, 
and inflation ends when $\Phi$ becomes lower than $M_P/\sqrt{\xi}$.

 The various subsequent stages in the evolution of $\Phi$ in the universe leading to leptogenesis are as follows~\cite{Stubbs, nobu}:
 
 \begin{itemize} 
 
\item  In the very early universe when $|\Phi| \gtrsim M_P/\sqrt{\xi}$, the non-minimal coupling in the Einstein frame leads to a constant $V^E(\Phi)$, and drives inflation, as stated above.

\item In the second phase, as the field  $|\Phi|$ has rolled sufficiently down the potential to its value less than $M_P/\sqrt{\xi}$, the effect of the non-minimal coupling becomes unimportant and inflation ends. 
The value of $|\Phi|$ is still large 
and the dominant term in the potential driving the evolution of the $|\Phi|$ is the $\lambda |\Phi|^4$ term.  
At the beginning of this stage, the real and imaginary parts of the field are already different, owing to the $\epsilon$ term or some other mechanism.
This asymmetry survives the evolution of the $\Phi$ field and eventually leads to the baryon asymmetry of the universe. This is the key idea in AD baryogenesis.

\item As the universe evolves further, $\Phi$ becomes smaller since it scales like $a^{-1}$ ($a$ being the scale factor) and the third stage begins where the quadratic term in the potential dominates over the quartic term for a suitable choice of the $\Phi$ mass. This leads to an oscillatory behavior of $|\Phi|$ and the universe behaves like it is matter dominated. This approximation of transition of the potential from being quartic dominated to quadratic dominated (when $|\Phi| \leq M_\Phi/\sqrt{\lambda}$) is called the threshold approximation in \cite{Stubbs}.

\item The fourth  stage is when the AD field decays to $ NN, N'N'$ states (more precisely, $N_1 N_1$, $N_1^\prime N_1^\prime$ as discussed in the previous section) and the $N$ and $N'$-fields decay immediately to standard model particles, if $\Gamma_{\Phi\to N, N'} < \Gamma_{N, N'}$. 
At this stage, the $\Phi$ asymmetry gets transferred to $N$ and $N'$ asymmetry. In the formulation given in Ref.~\cite{Stubbs, nobu}, the decay products are assumed to quickly thermalize, and the SM baryon asymmetry is generated due to sphaleron interactions.
\end{itemize}

There are now two possibilities for the particle that becomes the dark matter depending on whether  $v' $  is larger or smaller than the mirror reheating temperature $T'_R$.

\begin{enumerate}

 \item If in the mirror sector, $v' > T'_R$, which can happen for some choice of parameters, the mirror sphalerons would have gone out of equilibrium by the time mirror lepton asymmetry is generated. The mirror lepton asymmetry therefore does not get a chance to become mirror baryons. This lepton asymmetry resides in the lightest of the mirror leptons, the mirror electron, which then constitutes the dark matter of the universe. 
 
 \item On the other hand, if $v' < T'_R$, the mirror lepton asymmetry can get partially converted ($\frac{28}{79}$ of it) to mirror baryons. In this case, we have a combination of mirror baryons and/or mirror leptons become the dark matter depending on the mirror weak vev magnitude. We discuss the details below.
 
 \end{enumerate}

Case (1): Due to $Z_2$ symmetry of the model, the lepton asymmetries in the mirror and visible sectors are in the ratio $Br(\Phi\to N'N')\equiv Br$. Since $\Omega_{DM}\simeq \frac{n_L' m_{e'}}{\rho_c}$, in case (1) and $\Omega_B\simeq\frac{28}{79} \frac{n_L m_p }{\rho_c}$, demanding that $\Omega_{DM}= 5 \Omega_B$  implies that $m_{e'}\simeq 1.66/Br$ GeV. This explains the matter-dark matter coincidence puzzle. 
Now using the fact that $m_{e'}/m_e=v'/v=3.25\times 10^3/Br$,  we get $v'\simeq  5.66\times 10^5/Br$ GeV.

Case (2): Here $v' < T'_R$, as a result, part of the lepton asymmetry gets converted to mirror baryons. $\Omega_{DM}$ in this case consists partially of $\Omega_{e'}$ and the rest $\Omega_{B'}$. Using $\frac{\Omega_{DM}}{\Omega_B}=5$  we get 
\begin{eqnarray}
m_{B'}/ m_B\simeq 5/Br.
\end{eqnarray}
Relating this ratio to $v'/v$ is a bit involved since the mirror nucleon mass in a QCD like theory depends not only on the mirror quark masses but also on the $\Lambda'_{QCD}$. For $Br\simeq 0.01$ or so, the mirror baryon mass required for it to become DM is about 469 GeV. This is much larger than the  $\Lambda'_{QCD}$ (see Fig. 1). We therefore assume that the mirror quark masses provide the bulk of the mirror baryon mass. If $B'=p'$, then the three masses $u'u'd'$ add up to $m_{B^\prime} \simeq 2m_{u'}+m_{d'}= 4.69$ GeV$/Br$. 
Using simple scaling we expect $v'/v= m_{B^\prime}/(2 m_u+m_d)= [4.69~{\rm GeV}]/[20~{\rm MeV} Br]\simeq 235/Br$. 
This value of $v'/v$ is about an order of magnitude smaller than that required to make  $e'$ a dominant component of dark matter and therefore, 
$e'$ contribution to dark matter in this case is much smaller than the mirror baryon contribution. 

This is a key new result of the paper that we believe explains why the baryon asymmetry and DM abundance are of the same order~\footnote{A different argument based on strong CP resolution in mirror models also leads to mirror electron as the dark matter~\cite{hall, hall2}}. We note that to relate the $\Omega_M$ with $\Omega_{DM}$, it is important that dark 
matter is asymmetric type. This means that the symmetric part of the dark matter particle be negligible. We show below that this is indeed true.

\section{6. Quantitative aspects of dark matter generation}  
In order to get the main result of our paper, the model parameters must satisfy certain constraints, which we outline in  this section. We start with the formula for lepton asymmetry generated by the above
AD mechanism from references~\cite{Stubbs, nobu}. It is given in terms of the $L$-violating coefficient $\epsilon$ in Eq. (1), the reheat temperature in the visible sector $T_R$ as follows:
\begin{eqnarray}
~\frac{n_{L'}}{s}~=\frac{n_L Br }{s}~=~ \frac{T^3_R}{\epsilon M^2_\Phi M_P} Br.
\label{nL}
\end{eqnarray}
In this equation, $\epsilon$ must satisfy the condition that $1 \gg \epsilon \gg \frac{\Gamma_\Phi}{M_\Phi}$ \cite{Stubbs, nobu}. 
We must make sure to satisfy this relation, when we choose our benchmark values for model parameters. 
Furthermore, this will imply that  the branching ratio $Br(\Phi\to N'_1 N'_1)$ is related to $\Delta N_{eff}$, since the resulting mirror neutrinos appear as dark radiation. We discuss this below.

Let us define a parameter $K=T_R/M_\Phi$. For the model to work, we choose $K < 1$ so that by the time reheat temperature is reached, $\Phi$ is non-relativistic 
and its number density is Boltzmann suppressed. Hence, the inverse decays are Boltzmann suppressed, and washing-out process is inactive. 
We can also check that a washing-out processes mediated by $\Phi$ is out of equilibrium. 
We have $n_L/s =\frac{79}{28} \frac{n_B}{s}\simeq  2.5\times 10^{-10}$ (using $\frac{n_B}{s}=8.7\times 10^{-11}$), Eq.~(\ref{nL}) implies
\begin{equation}
M_{\Phi}\simeq  6.0\times 10^{8} \frac{\epsilon}{K^3} ~{\rm GeV},
\label{eq1}
\end{equation}
Alternatively, we can write the same constraint as
\begin{eqnarray}
\epsilon~=\left(\frac{79}{28} \frac{n_B}{s}\right)^{-1}  \left( \frac{M_\Phi}{M_P} \right) K^3.
\label{ep1}
\end{eqnarray}

Using the sudden decay approximation, we estimate the reheating temperature by 
$\Gamma_\Phi=\frac{y^2_\Phi}{4\pi} M_\Phi=H=\left(\frac{\pi^2 g_*}{90}\right)^{1/2}\frac{T^2_R}{M_P}$. 
Using $T_R=KM_\Phi$, Eq.~(\ref{eq1}) and $g_*\simeq 106.75$, we find the following constraint:
\begin{eqnarray}
\frac{\epsilon}{K}~=~9.48 \times 10^7 \, y^2_\Phi.
\label{ep2}
\end{eqnarray}

As alluded to in sec.~5, the SM lepton asymmetry $n_L$ is converted to baryon asymmetry via the sphaleron effect. 
For this conversion to be active, 
we must have $T_R > T_{sph} \simeq 130$ GeV, where $T_{sph}$ is the sphaleron decoupling temperature. 
Using $T_R=K M_{\Phi}$ and Eq.~(\ref{eq1}), the $T_R$ lower limit leads to
\begin{eqnarray}
\frac{\epsilon}{K^2} > 2.18\times 10^{-7} .
\label{ep3}
\end{eqnarray}

For consistency of the $n_L/s$ formula, we must have $1 \gg \epsilon \gg \Gamma_\Phi/M_\Phi$, which leads to
\begin{eqnarray}
\epsilon \gg y^2_\Phi/4\pi .
\label{ep4}
\end{eqnarray}
Setting this condition, for simplicity, to be $\epsilon > 10  \times y^2_\Phi/4\pi$, and using Eq.~(\ref{ep2}), we get
\begin{eqnarray}
K > 8.4\times 10^{-9} .
\label{K1}
\end{eqnarray}
Note that this is a rather weak bound.



\begin{figure}[tb]
  \centering
 \includegraphics[width=0.7\linewidth]{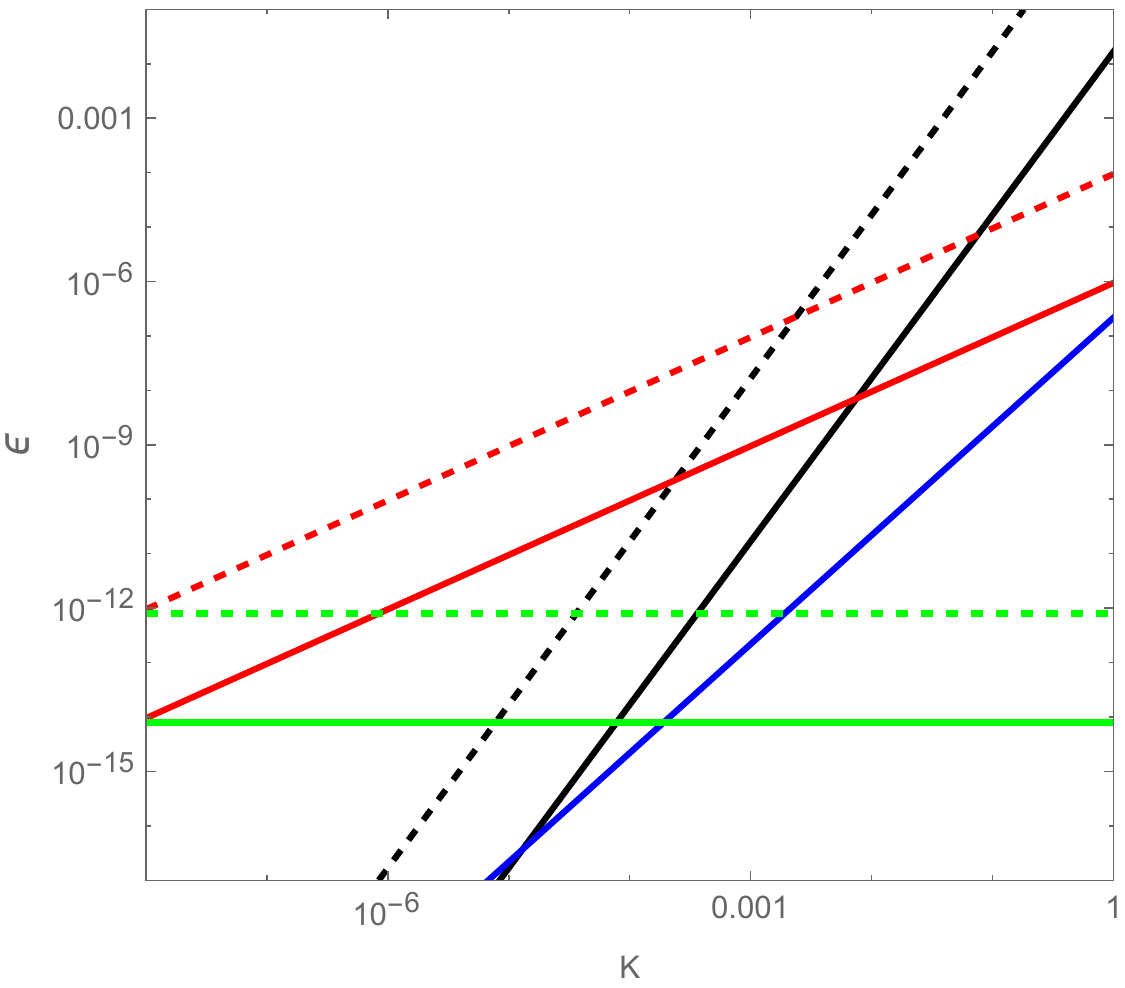}
  \caption{
Restriction on the model parameters $K$ and $\epsilon$ from 
Eqs.~(\ref{ep1}), (\ref{ep2}), (\ref{ep3}), (\ref{ep4}) and (\ref{K1}).   
We show plots for two benchmarks sets on Table II: 
$(M_\Phi, y_\Phi)=(10^6~{\rm GeV}, 10^{-7})$ for set I, and $(M_\Phi, y_\Phi)=(10^{12}~{\rm GeV}, 10^{-5})$ for set II. 
For set I, Eqs.~(\ref{ep1}) and (\ref{ep2}) are depicted as 
The black and red solid lines, respectively, represent Eqs.~(\ref{ep1}) and (\ref{ep2}) for set I. 
The black and red dashed lines corresponds to set II. 
Allowed values of $K$ and $\epsilon$ are determined by the intersection of these lines. 
The blue diagonal line represents Eq.~(\ref{ep3}), and the region to the left of this line is allowed. 
The green horizontal lines represent $\epsilon = 10  \times y^2_\Phi/4\pi$ for set I (solid) and set II (dashed), respectively, 
and the regions above the lines are allowed. 
%
 } 
  \label{fig:2}
  \end{figure}

If we choose $M_\Phi$ and $y_\Phi$ as our primary free parameters, 
$\epsilon$ and $K$ are determined as a solution which simultaneously satisfies Eqs.~(\ref{ep1}) and (\ref{ep2}). 
The values of $\epsilon$ and $K$ determined in this way must satisfy Eqs.~(\ref{ep3}), (\ref{ep4}) and (\ref{K1}). 
In Fig.~\ref{fig:2}, we show plots for two benchmarks choices (see Table II) for Set I: $(M_\Phi, y_\Phi)=(10^6~{\rm GeV}, 10^{-7})$  
and Set II: $(M_\Phi, y_\Phi)=(10^{12}~{\rm GeV}, 10^{-5})$. 
For Set I, Eqs.~(\ref{ep1}) and (\ref{ep2}) are depicted as the black and red solid lines, respectively. 
$K$ and $\epsilon$ are determined by the intersection of these two lines. 
The black and red dashed lines corresponds to Set II. 
The blue diagonal line represents Eq.~(\ref{ep3}), and the region to the left of this line is allowed. 
The green horizontal lines represent $\epsilon = 10  \times y^2_\Phi/4\pi$ for Set I (solid) and Set II (dashed), respectively, 
and the region above the lines are allowed.

\begin{table}
\begin{center}
\begin{tabular}{|c||c||c|}\hline
Parameters & Set I& Set II\\\hline
$M_\Phi$  & $ 10^{7}~{\rm GeV}$& $10^{10}~{\rm GeV}$\\ 
$y_\Phi$ &$10^{-7}$ &$10^{-6}$\\
\hline
$K$ & $7.5 \times 10^{-3}$ & $2.4 \times 10^{-3}$\\
$\epsilon$ & $ 7.1 \times 10^{-9}$  & $ 2.3 \times 10^{-7}$\\ 
\hline
$M_{N}$ & $10^{6}$ GeV & $10^{9}$ GeV \\
$y_N$ & $1.3 \times 10^{-7}$ &$ 1.5 \times 10^{-5}$ \\
$v'$ & $ 5.7 \times 10^7$ GeV &$ 4.1 \times 10^6$ GeV\\
$T_{R}$ & $ 7.5 \times 10^4 $ GeV&$ 2.4 \times 10^{7}$ GeV\\\hline
\end{tabular}
\end{center}
\caption{ 
Benchmark sets of parameters for the model to illustrate that the model works. For the first set choice of parameters, $v' > T_R$ and therefore mirror  electrons are dark matter. In the second set $v' <T'_R$ so that mirror protons are dark matter. 
We have set $Br=0.01$, and $y_N$ is chosen to get $m_\nu = 5\times 10^{-11}$ GeV. 
}
\end{table}

\section{7. Other constraints} 
There are several other cosmological constraints on the model and we will now check if they are consistent with the allowed parameter region shown in Fig.~\ref{fig:2}. 

\subsection{7.1 Lepton number washout} 
The process dominantly contributing to wash-out of lepton asymmetry in the model is 
$NN\to \bar{N}\bar{N}$ and $N'N'\to \bar{N'} \bar{N'}$. 
 This is tree-level process mediated by $\Phi$ exchange whose rate at $T = M_\Phi$ is given by
 \begin{equation}
 \langle \sigma_{NN\to \bar{N}\bar{N}}\cdot v \rangle \simeq \frac{y^4_\Phi \epsilon^2 }{4\pi M^2_\Phi}. 
 \end{equation}
To avoid this wash-out, we demand that 
 $n_N \langle \sigma_{NN\to \bar{N}\bar{N}}v \rangle  <  H(T=M_\Phi) = 
 \left( \frac{\pi^2}{90} g_* \right)^{1/2} \frac{M_\Phi^2}{M_P}$,  
 where $n_N \simeq \frac{3}{2 \pi^2} M_\Phi^3$ is the number density of $N$.  
This means that the wash-out process is out-of-equilibrium, leading to  
 \begin{equation}
 M_\Phi > 3.5 \times 10^{-3} \, y_\Phi^{4} \, \epsilon^2 \, M_P ,
 \end{equation}
 This implies that
 \begin{eqnarray}
 y^4_\Phi < \frac{7.1 \times 10^{-8}}{\epsilon K^3} .
 \label{phi1}
 \end{eqnarray} 
 As we have discussed above, we set $\epsilon \ll 1$ and $ K < 1$, 
 so that $y_\Phi \lesssim 0.01$ always satisfies this bound. 
 
 
 

\subsection{7.2 Depletion of the symmetric part  of $B'$ and $e'$  content} 
For mirror electrons to be the asymmetric dark matter, the symmetric parts of it i.e.  abundance of  $e^{\prime -}$ and $e^{\prime +}$ and similarly $q'$ and $\bar{q'}$ must be very small. This requires us to find out when $q'\bar{q'} \to G'G'$ and $e^{\prime -}e^{\prime +}\to \gamma'\gamma'$ processes decouples from the thermal bath. 

Let us first discuss the mirror quark case. 
In the non-relativistic regime, the annihilation cross-section of mirror quarks is given by
\begin{equation}
\langle \sigma_{q'\bar{q}'\to G'G'}\cdot v \rangle \simeq \frac{4\pi (\alpha^\prime_s)^2}{m^2_{q'}} . 
\label{cross-sec}
\end{equation} 
We determine the decoupling temperature of the process by setting the mirror quark annihilation rate to the Hubble rate: 
\begin{equation}
n_{q^\prime}(T_d^\prime) \, 
\langle \sigma_{q'\bar{q}'\to G'G'}\cdot v \rangle
= \sqrt{\frac{\pi^2 g_*}{90}}\frac{T^{2}_d}{M_P} ,
\label{Td}
\end{equation}
where $n_{q^\prime}(T_d^\prime)$ is the number density of the mirror quark at the decoupling 
approximately given by  
\begin{eqnarray}
n_{q^\prime}(T_d^\prime) \simeq 4 N_c \left( \frac{m_{q'}T_d^\prime }{2\pi} \right)^{3/2} 
e^{-\frac{m_{q'}}{T^\prime_d}} 
\label{neq}
\end{eqnarray}
in the non-relativistic limit. 
Note that the decoupling temperatures in the SM sector ($T_d$) and in the mirror sector ($T_d^\prime$) 
are related by the formula $T_d^\prime=Br^{1/4} \, T_d$, and $m_{q^\prime}= m_q  (v^\prime/v)$. 
As discussed in sec.~6, if the mirror proton is the dark matter, $v'/v=235/Br$, and hence 
$m_{u^\prime}=1.18/Br$ GeV. 
For a fixed value of $Br$, we numerically solve this equation to find $T_d^\prime$ with $\alpha^\prime_s=0.12$. 
Then, the yield of the mirror u-quark at the decoupling is estimated by 
\begin{eqnarray}
Y_{u^\prime} (T_d^\prime) \simeq  \frac{n_{q^\prime}(T_d^\prime)}{s(T_d)}, 
\end{eqnarray}
where $s(T_d)$ is the entropy density of the universe, which is approximately the entropy density of 
the SM sector: 
\begin{eqnarray}
s(T_d) = \frac{2 \pi^2}{45} g_* T_d^3 =  \frac{2 \pi^2}{45} g_* \left( \frac{T_d^\prime}{Br^{1/4}} \right)^3. 
\label{etp}
\end{eqnarray}
The relic density of the mirror u-quark is estimated by 
\begin{equation}
 \Omega_{u'}h^2\simeq \frac{ m_{u^\prime} \, Y_{u^\prime}(T_d^\prime)  \, s_0}{\rho_c /h^2},  
\end{equation}
where $s_0=2890$/$cm^3$ is the entropy density of the present universe, 
and $\rho_c /h^2=1.05 \times 10^{-5}$ GeV/$cm^3$ is the critical density. 
For $Br=0.01$, we find $T_d^\prime =3.5$ GeV and $\Omega_{u'}h^2 =6.8 \times 10^{-6}$, 
which is much smaller than the dark matter density of $\Omega_{DM}h^2 =0.12$. 

 \begin{figure}[tb]
  \centering
 \includegraphics[width=0.42\linewidth]{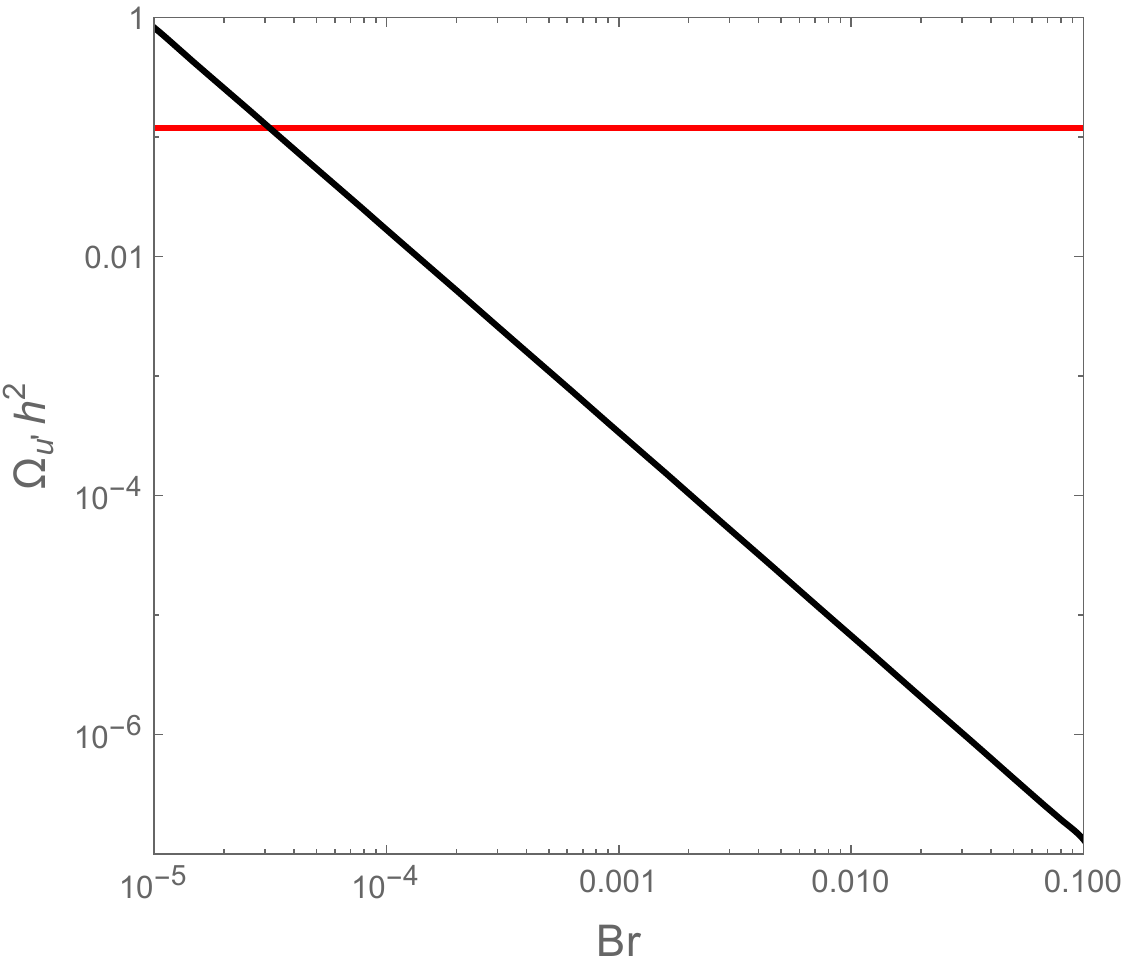} ~~~~
 \includegraphics[width=0.42\linewidth]{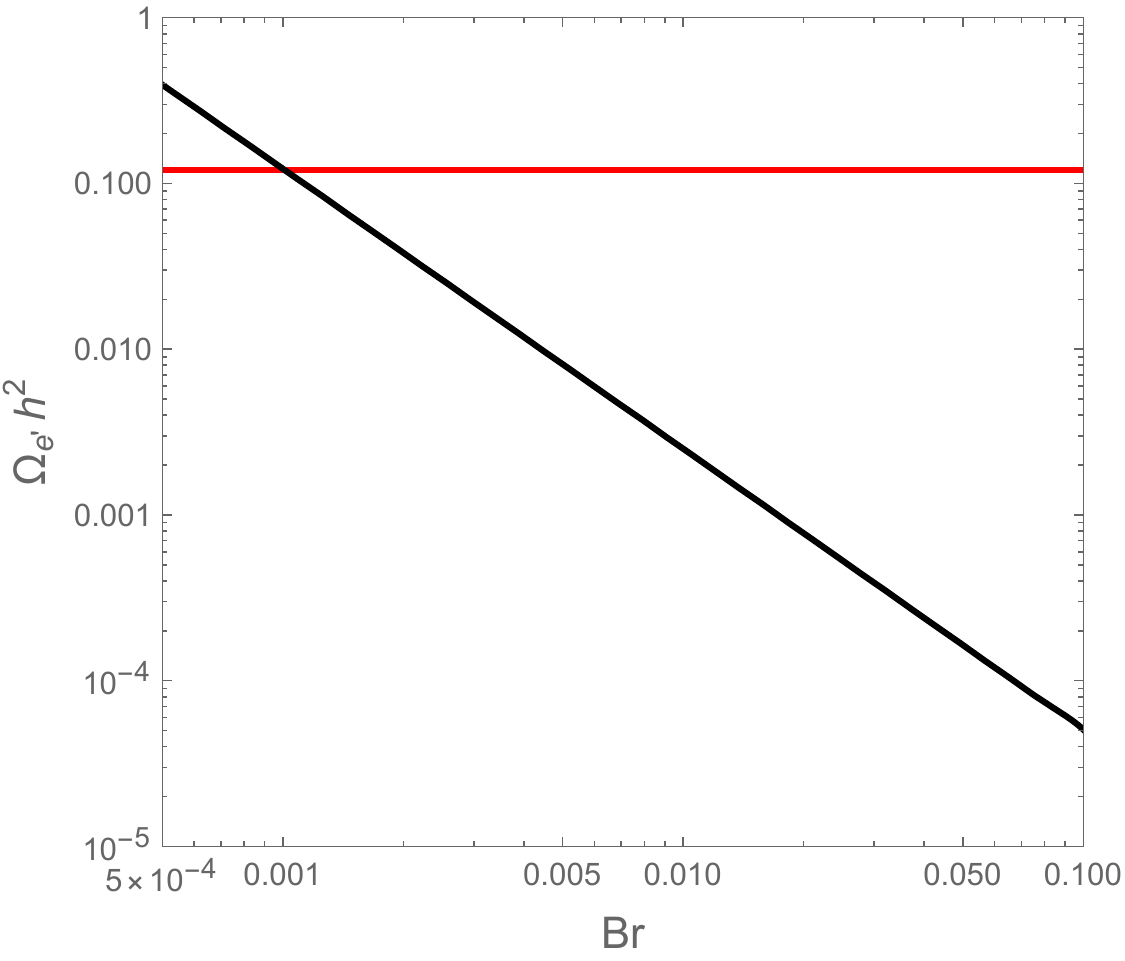} 
  \caption{
The black lines represent the relic abundances of the symmetric parts for the case that the mirror proton 
is the asymmetric dark matter  in the left panel, and for the case that the mirror electron is the asymmetric dark matter  in
the right panel.  The symmetric parts of the relic abundances, in both cases are proportional to $Br^{-7/4}$. 
The horizontal red lines indicate the observed dark matter density, $\Omega h^2 =0.12$.  These graphs lead to the lower limit on the branching ratio $Br$ from the requirement that the dark matter is asymmetric.  This in turn gives the lower limit
on dark radiation parameter $\Delta N_{eff}$.
} 
  \label{fig:3}
  \end{figure}

We can see that the resultant $\Omega_{u'}h^2$ increases as we lower $Br$. 
This behavior can be understood as follows. 
From Eq.~(\ref{Td}) and (\ref{etp}), 
\begin{eqnarray}
Y_{u^\prime} (T_d^\prime) &=&
 \frac{n_{q^\prime}(T_d^\prime)}{s(T_d)}
=\sqrt{\frac{\pi^2 g_*}{90}}\frac{T^{2}_d}{M_P} 
\langle \sigma_{q'\bar{q}'\to G'G'}\cdot v \rangle^{-1} 
\left(\frac{2 \pi^2}{45} g_* \left( \frac{T_d^\prime}{Br^{1/4}} \right)^3 \right)^{-1} \nonumber\\
&=&
\sqrt{\frac{45}{8 \pi^2 g_*}} \frac{(1.18\, {\rm GeV})^2}{4 \pi \,(\alpha^\prime_s)^2 \, M_P \, T_d^\prime} Br^{-7/4}. 
\end{eqnarray}
Hence, we find $\Omega_{u'}h^2 \simeq 6.4 \times 10^{-11} \, x_d^\prime/ Br^{7/4}$, 
where $x_d^\prime \equiv m_{u^\prime}/T_d^\prime$. 
Our calculation is essentially the same as the calculation for the relic density of the Weakly Interacting Massive Particle dark matter, and as is well known, $x_d^\prime =20-30$ over the wide range of the dark matter mass. 
In our numerical analysis, we find $23.3 \leq x_d^\prime \leq 37.3 $ for $10^{-5} \leq  Br \leq 0.1$.  
In the left panel of Fig.~\ref{fig:3}, we show our numerical result for $\Omega_{u'}h^2$ as a function of $Br$, 
where we can see $\Omega_{u'}h^2 \propto Br^{-7/4}$. 
We have set the model parameters to reproduce the observed dark matter density 
by the asymmetric dark matter and therefore, the relic density for the symmetric parts must be 
$\Omega_{u'}h^2 \ll 0.12$, which lead to the lower bound on $Br > 3 \times 10^{-5}$. 

 
For mirror electrons to be the asymmetric dark matter, we can repeat  the same calculation 
by replacing $\alpha'_s$ to $\alpha'_{em}=1/128$, 
$m_{u^\prime}=1.18/Br$ GeV to $m_{e'}= 1.66/Br$ GeV, 
and $n_{u^\prime}$ to $n_{e^\prime}$ by setting $N_c=3 \to 1$ in Eq.~(\ref{neq}). 
Our result is shown in the right panel of Fig.~\ref{fig:3}. 
For this case, we find the lower bound $Br > 10^{-3}$. 

We also note that if the asymmetric part provides the central value of $\Omega_{DM}$, then the symmetric part can at most be within $2\sigma$ uncertainty of this which from 
~\cite{Planck:2018vyg} is $\simeq .0024$ at 95\% confidence level. This would imply from ~Fig. \ref{fig:4} that $Br \geq 0.01$.

\subsection{7.3 Dark radiation and $\Delta N_{eff}$ } 
In this model, the mirror neutrinos and the mirror photon, if it is light enough, remain as dark radiation contributing to $\Delta N_{eff}=(N_{eff}-N^{SM}_{eff})$. We now evaluate $\Delta N_{eff}$ using the same technique as in \cite{chacko2}. Let us denote ${g}_{*R}$ and ${g}'_{*R}$ as the number of degrees of freedom in the visible and mirror sectors, respectively, at the moment right after the reheat epoch when the temperatures in the visible and the mirror sectors are ${T_R}$ and ${T'_R}$. If the corresponding number of degrees of freedom at the BBN epoch are  $g_{*BBN}$ and $g'_{*BBN}$ in the two sectors and the temperatures at the BBN epoch be $T_{BBN}$ and $T'_{BBN}$, we can use entropy conservation to get the following relations among the above temperatures and degrees of freedom parameters.   Entropy conservation in both sectors then leads to:
 \begin{eqnarray}
 \frac{{g}'_{*R} {T'_R}^{3}}{g_{*R}{T_R}^3}=\frac{g'_{*BBN}{T' _{BBN}}^3}{g_{*BBN}{T_{BBN}}^3}
 \end{eqnarray}
 Since the numerator and the denominators on each side of the above equation are at the same cosmic time, the volume factors cancel out.
 From the asymmetric reheating, we get the relation between  ${T_R}$ and ${T_R'}$ to be
 \begin{eqnarray}
 {g}'_{*R} {T_R'}^4 =Br \times  {g}_{*R} {T_R}^4 .
 \end{eqnarray}
 Using the definition of $\Delta N_{eff}$, these two equations give
 \begin{eqnarray}
 \Delta N_{eff}~=~\frac{{g'}_{* BBN}T^{'4}_{BBN}}{\frac{7}{4} T^4_{BBN}}=\frac{4 Br}{7} g_{* BBN}\left(\frac{{g}'_{*R}\cdot g_{*BBN}}{{g}_{*R}\cdot {g'}_{*BBN} }\right)^{1/3}
 \end{eqnarray}
 Thus, we relate the amount of dark radiation $\Delta N_{eff}$ to the unknown parameter $Br$. 
 If we make a plausible assumption that  ${g}_{*R}={g}'_{*R}$, we get
  \begin{eqnarray}
 \Delta N_{eff}~=~\frac{4 Br}{7} g_{* BBN}\left(\frac{g_{*BBN}}{g'_{*BBN} }\right)^{1/3} .
 \end{eqnarray}
 If at the epoch of BBN, only mirror neutrinos and the mirror photon contribute to $g'_*$, 
 then we get  $g'_{*BBN}=7.25$. Using the fact that in the visible sector $g_{*BBN}=10.75$, we get
 $\Delta N_{eff}\simeq 7.0 \times Br$. The current bound on  $\Delta N_{eff} < 0.284$ at 95\% C.L. 
 from Planck data~\cite{Planck:2018vyg} then implies $Br <  0.041$. This leads to the cooling factor of the mirror world as $x\equiv \frac{T^\prime_R}{T_R}= 0.45 \simeq \delta_\beta$. This is enough to guarantee the success 
of the standard BBN results in a mirror world picture. 
The case with the mirror electron being the asymmetric dark matter is especially interesting, since choosing the symmetric part to be a tenth of the observed relic density gives a lower 
bound on $Br > 10^{-3}$ as shown in Fig.~\ref{fig:3}. 
This translates to a lower bound on $\Delta N_{eff} > 7.0 \times 10^{-3}$, part of which region will be explored 
by future precision CMB experiments such as CMB-S4 \cite{CMB-S4:2016ple} and CMB-HD~\cite{HD}. If on the other hand we assume that the uncertainty in $\Omega_{DM}=0.1200\pm 0.0024$ (95\% c.l.) is due to the symmetric part of the dark matter, then we get a lower limit on $Br \geq 0.01$ leading to a lower bound of  $\Delta N_{eff} > 0.07$, which is testable in CMB-S4 and CMB-HD experiments.

\subsection{7.4 Preventing Cross-equilibrium from Dirac masses of singlet fermions}

\begin{figure}[tb]
  \centering
 \includegraphics[width=0.6\linewidth]{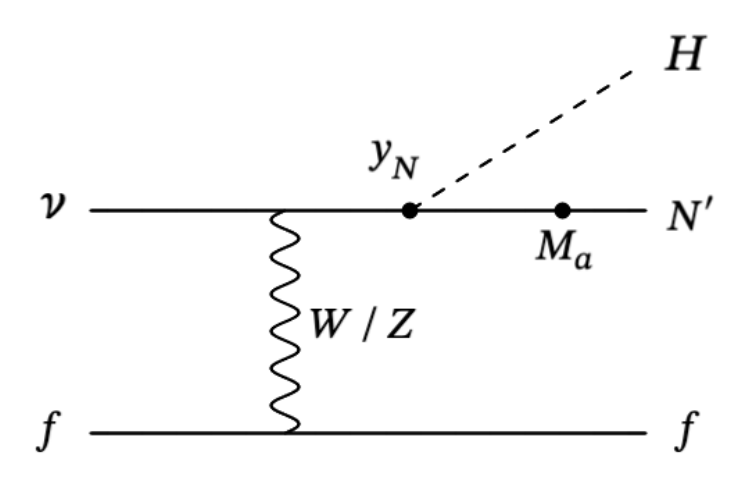}  
  \caption{
The Feynman diagram for the process  $f+\nu \leftrightarrow  H+N^\prime+f$ mediated by $W$ and $Z$ bosons, that connects the SM with mirror sector via the Dirac neutrino portal.
} 
  \label{fig:4}
  \end{figure}

We see from our discussion of neutrino masses in sec.~3 that the Dirac masses of the singlet fermions $N, N'$ and $S, S'$ connect the visible sector to the mirror sector via their large masses $M_{1, 2, 3, 4}$. 
They also mix with the familiar and mirror neutrinos via Yukawa couplings. Since the latter have gauge interactions, there are processes like $f+\nu \leftrightarrow  H+N^\prime+f$ mediated by $W$ and $Z$ bosons (see Fig.~\ref{fig:4}), 
where $f$ are SM fermions (quarks and leptons). 
Once this process gets into thermal equilibrium, it will establish cross-equilibrium between the visible and mirror sector undoing the effect of asymmetric reheating and create problems for BBN. This process must therefore be out of equilibrium. To study this, note that the out of equilibrium condition for this process at temperature $T\gg v$ is given by
\begin{eqnarray}
n_f  \langle \sigma (f+\nu \leftrightarrow H+N^\prime+f) \cdot v \rangle \sim 0.1\, T^3 \,  \frac{y^2_N g_2^4 M^2_a}{8\pi^3 T^4} <  \frac{T^2}{M_P}, 
\end{eqnarray}
where $g_2$ is the SM $SU(2)$ gauge coupling.  
 This leads to the decoupling $T_*$ for this process to be 
\begin{eqnarray}
T_*\simeq \left( \frac{0.1 \, y^2_N \, g_2^4 \,  M^2_N M_P}{8\pi^3}\right)^{1/3} , 
\end{eqnarray}
where we have set $M_a=M_N$. 
As an example, if we consider the benchmark set I and II in Table II, we get 
$T_* \sim 10^4$ GeV and $T_* \sim 10^7$ GeV, respectively, 
above which the process is out of equilibrium. 
For both cases, $T_*$ is much smaller than $M_N$, and therefore, 
the processes are Boltzmann suppressed by $e^{-M_N/T}$ for $T < T_*$ 
and automatically keeping the process out of equilibrium.  In a similar manner, the process $f'+\nu' \leftrightarrow N+f'$ decouples above the 
reheat temperature $T_R$ and prevents cross-equilibrium from the mirror world to the SM sector.

\section {8. Self-interacting Dark Matter (SIDM) and massive mirror photon} 
The considerations of this paper implies that the lepton excess in the mirror sector is created below the temperature,  where mirror sphalerons decouple. As a result, the mirror lepton excess remains as it is and therefore constitutes  the dark matter content of the universe. 
As we have discussed in sec.~6, we see that 
\begin{equation}
\frac{n_{L'}}{s}~\simeq  \frac{79}{28} \frac{n_B}{s}Br
\end{equation}
due to SM sphaleron effect. This implies that mirror lepton excess gives the correct dark matter content if $m_{e'}\simeq 1.66/Br$ GeV as noted above. However, not only do the mirror leptons experience the mirror Coulomb force among them but that is also a repulsive force. We therefore need to make sure that this is in agreement with current structure observations. We use the detailed work of ref.~\cite{haibo2} for this purpose, where this issue is addressed. They point out that both attractive and repulsive forces can fit structure observation data as long as the force has finite range (see Fig. 6 of ref.~\cite{haibo2}). In terms of our mirror model, it means that the mirror photon must have mass. Their Fig.~6 also informs us that, for a few GeV mirror electron mass, the mirror photon mass must be in the range of 10-100 MeV. The bullet cluster bound~\cite{bullet} on SIDM is $\frac{\sigma_{DM-DM}}{m_{DM}}\leq 1 {\rm cm}^2/ g\simeq 1.8\times 10^{-24}  {\rm cm}^2$/GeV. We estimate the tree-level $e'e'$ scattering cross section to be $\sigma_{e'e'}\simeq \frac{4\pi \alpha^2_{em} }{m^2_{e'}}\sim 10^{-26}  {\rm cm}^2$, for $m_{\gamma'}\leq 100$ MeV in agreement with the above bound. 

In our model, the mirror photon mass arises from the non-zero vev of $\eta'$ scalar. One might think that from the coupling $\ell'_e\ell'_j \eta^{'}$, there will result a coupling of the form $e' (Re \eta') \nu'$ after $\eta'$ vev  The mass of the $(Re~\eta')$ however can be made $\sim v'$ which will prevent $e'$ dark matter decay.

\section{9. Conclusion} In conclusion, we have shown how the mirror universe model can provide an explanation of why the dark matter content of the universe is of the same order of magnitude as the baryon excess. The model guarantees the needed temperature asymmetry between the two sectors, using the neutrino mass mechanism in the model. A unique feature of our model is that a single complex field ($\Phi$) is responsible for inflation, asymmetric reheating, baryogenesis as well as solving the baryon-dark matter coincidence problem. We find this unified explanation of several problems of the standard model to be quite appealing. The neutrinos in the model are Dirac fermions. A prediction of the model is the lower bound on the amount of dark radiation ($\Delta N_{eff}$) in the universe. The forth coming CMBS4 as well as the CMB-HD experiment can probe a large range of the $\Delta N_{eff}$ prediction.


\section*{Acknowledgement}

We thank Hai-bo Yu for very helpful discussions on self-interacting dark matter and for reading an earlier version of the manuscript. 
The work of N. O. is supported in part by the United States Department of Energy Grant DE-SC0012447 and DE-SC0023713.


\begin{thebibliography}{99}

\bibitem{LY} T.D. Lee and C.N. Yang, Phys. Rev. {\bf 104}, 254 (1956).

\bibitem{KOZ} I.Yu. Kobzarev, L.B. Okun and I.Ya. Pomeranchuk, Yad. Fiz.{\bf  3}, 1154 (1966), [Sov. J.
Nucl. Phys. {\bf 3}, 837 (1966)].

\bibitem{M1} M. Pavsic, Int. J. Theor. Phys. {\bf 9}, 229 (1974).

\bibitem{M2} L.B. Okun, JETP {\bf 79}, 694 (1980).

\bibitem{M3}  S. Blinnikov and M. Khlopov, Sov. Astron. 27, 371 (1983).

\bibitem{M4}  R. Foot, H. Lew and R.R. Volkas, Phys. Lett. B{\bf 272}, 67 (1991). 12. and Mod. Phys. Lett. {\bf A7}, 2567 (1992);

\bibitem{M5} R. Foot and R. Volkas, Phys. Rev. {\bf D 52}, 6595 (1995).

\bibitem{M6} Z. Berezhiani and R.N. Mohapatra, Phys. Rev. {\bf D 52}, 6607 (1995).

\bibitem{M7} Z. Chacko, H.-S. Goh, and R. Harnik, Phys. Rev. Lett. {\bf 96}, 231802 (2006), arXiv:hep-ph/0506256.

\bibitem{Kolb} 
E.W. Kolb, D. Seckel, M.S. Turner, Nature {\bf 314} (1985) 415-419.

\bibitem{Hodges} H.M. Hodges, Phys.Rev. {\bf D 47} (1993) 456-459.

\bibitem{BDM} Z.G. Berezhiani, A.D. Dolgov, R.N. Mohapatra, Phys. Lett. {\bf B 375} (1996) 26-36.

\bibitem{Vilenkin} V.S. Berezinsky, A. Vilenkin,  Phys.Rev. {\bf D 62} (2000) 083512, e-Print: hep-ph/9908257 [hep-ph].

\bibitem{cline} 
J.M. Cline, J.S. Roux, Phys.Rev. {\bf D 105} (2022) 4, 043506, e-Print: 2107.04045 [astro-ph.CO].

\bibitem{chacko} 
Z. Chacko, N. Craig, P.J. Fox, R. Harnik, JHEP {\bf 07} (2017) 023, e-Print: 1611.07975 [hep-ph].

\bibitem{shmuel} S. Nussinov,  Phys. Lett. B {\bf  165},  (1985) 55-58. 

\bibitem{sekhar} S. M. Barr, R. S. Chivukula, E. Farhi,  Phys. Lett. B, {\bf 241}, (1990) 387-391.

\bibitem{kaplan}  D. E. Kaplan, M. A. Luty, and K. M. Zurek, Phys. Rev. {\bf D 79}, 115016 (2009), arXiv:0901.4117 [hep-ph].

\bibitem{DM}  H. Davoudiasl and R. N. Mohapatra, New J. Phys. {\bf 14}, 095011 (2012), arXiv:1203.1247 [hep-ph].


\bibitem{PV}  K. Petraki and R. R. Volkas, Int. J. Mod. Phys. {\bf A 28}, 1330028 (2013), arXiv:1305.4939 [hep-ph].


\bibitem{zurek} K. M. Zurek, Phys. Rept. 537, 91 (2014), arXiv:1308.0338 [hep-ph].




\bibitem{An} 
H. An, S.L. Chen, R.N. Mohapatra, Y. Zhang, JHEP {\bf 03} (2010) 124.

\bibitem{cui}  Y. Cui and R. Sundrum, Phys. Rev. {\bf D 87}, 116013 (2013), arXiv:1212.2973 [hep-ph].

\bibitem{nath}  
W.Z. Feng, P. Nath; Phys. Lett. {\bf B 731} (2014) 43-50.

\bibitem{farina} M. Farina, JCAP  {\bf 11}, 017 (2015), arXiv:1506.03520 [hep-ph].

\bibitem{FMS} 
M. Farina, A. Monteux, C.S. Shin, Phys.Rev. {\bf D 94} (2016) 3, 035017.

\bibitem{Raman} 
A. Bodas, M.A. Buen-Abad, A. Hook, R. Sundrum, JHEP {\bf 06} (2024) 052. 


\bibitem{volkas}  S. J. Lonsdale and R. R. Volkas, Phys. Rev. D 90, 083501 (2014), [Erratum: Phys.Rev.D 91, 129906 (2015)],
arXiv:1407.4192 [hep-ph].

\bibitem{ibe} M. Ibe, A. Kamada, S. Kobayashi, T. Kuwahara, and W. Nakano, Phys. Rev. {\bf D 100}, 075022 (2019), arXiv:1907.03404
[hep-ph].

\bibitem{curtin} G. Alonso-Alvarez, D. Curtin, A. Rasovic, and Z. Yuan, (2023), arXiv:2311.06341 [hep-ph].

\bibitem{burdman} P. Bittar, G. Burdman, and L. Kiriliuk, JHEP  {\bf 11}, 043 (2023), arXiv:2307.04662 [hep-ph].


\bibitem{murgui} C. Murgui and K. M. Zurek, Phys. Rev. {\bf D 105}, 095002 (2022), arXiv:2112.08374 [hep-ph].


\bibitem{yi} Yi Chung, e-Print: 2411.16860 [hep-ph].

\bibitem{Hook} 
D. Brzeminski, A. Hook,  Phys.Rev.Lett. {\bf 132} (2024) 20, 201001.


\bibitem{Abhi}  
A. Banerjee, D. Brzeminski, A. Hook;  e-Print: 2410.22412 [hep-ph].


\bibitem{borah} 
D. Borah, S.J. Das, N. Okada, JHEP {\bf 05} (2023) 004, e-Print: 2212.04516 [hep-ph].


\bibitem{AD} 
I.~Affleck and M.~Dine,
Nucl. Phys. B \textbf{249}, 361-380 (1985). 



\bibitem{spergel}  
D.N. Spergel and P.J. Steinhardt, Phys. Rev. Lett. {\bf  84} (2000) 3760-3763, e-Print: astro-ph/9909386 [astro-ph]


\bibitem{haibo} 
S. Tulin and H.B. Yu, Phys. Rept. {\bf 730} (2018) 1-57 e-Print: 1705.02358 [hep-ph].


\bibitem{haibo2}  
S. Tulin, H.B. Yu, K.M. Zurek, Phys.Rev. {\bf D 87}, (2013) 11, 115007 e-Print: 1302.3898 [hep-ph].


\bibitem{Stubbs} 
A.~Lloyd-Stubbs and J.~McDonald,
Phys. Rev. D \textbf{103}, 123514 (2021).
[arXiv:2008.04339 [hep-ph]].

\bibitem{russian} 
E.~Babichev, D.~Gorbunov and S.~Ramazanov,
Phys. Lett. B \textbf{792}, 228-232 (2019).
[arXiv:1809.08108 [astro-ph.CO]].

\bibitem{cline2} J.~M.~Cline, M.~Puel and T.~Toma,
Phys. Rev. D \textbf{101}, no.4, 043014 (2020)  
[arXiv:1909.12300 [hep-ph]].


\bibitem{herzberg}  M.~P.~Hertzberg and J.~Karouby,
Phys. Lett. B \textbf{737}, 34-38 (2014) 
[arXiv:1309.0007 [hep-ph]]. 
%


\bibitem{nobu} 
R.~N.~Mohapatra and N.~Okada,
Phys. Rev. D \textbf{104}, no.5, 055030 (2021)
[arXiv:2107.01514 [hep-ph]].

\bibitem{Bezrukov}  F. L. Bezrukov and M. Shaposhnikov, Phys. Lett. {\bf B 659}, 703 (2008), [arXiv:0710.3755
[hep-th]].

\bibitem{hall}  
Q. Bonnefoy, L.J. Hall, C.A. Manzari, A. McCune, C. Scherb, Phys. Rev. D 109 (2024) 5, 055045, e-Print: 2311.00702 [hep-ph]


\bibitem{hall2}  
D.I. Dunsky, L.J. Hall, K. Harigaya, JHEP 02 (2024) 212, e-Print: 2302.04274 [hep-ph]


\bibitem{chacko2} Z. Chacko, Y. Cui, S. Hong and T. Okui,  Phys. Rev. {\bf D 92}, 055033 (2015).

\bibitem{Planck:2018vyg}
N.~Aghanim \textit{et al.} [Planck],
Astron. Astrophys. \textbf{641}, A6 (2020)
[erratum: Astron. Astrophys. \textbf{652}, C4 (2021)]
[arXiv:1807.06209 [astro-ph.CO]].


\bibitem{CMB-S4:2016ple}
K.~N.~Abazajian \textit{et al.} [CMB-S4],
[arXiv:1610.02743 [astro-ph.CO]].

\bibitem{HD} CMB-HD collaboration, S. Aiola \textit{et. al.}, arXiv:2203.05728.

\bibitem{bullet} 
M.~Markevitch, A.~H.~Gonzalez, D.~Clowe, A.~Vikhlinin, L.~David, W.~Forman, C.~Jones, S.~Murray and W.~Tucker,
Astrophys. J. \textbf{606} (2004), 819-824
[arXiv:astro-ph/0309303 [astro-ph]].


\end{thebibliography}
\end{document}